\documentclass[twocolumn]{aastex63}

\usepackage{amssymb}
\usepackage{amsmath}
\usepackage{xcolor}
\usepackage{verbatim}

\usepackage{booktabs}
\usepackage{natbib}

\submitjournal{ApJ}
\usepackage{lineno}

\shortauthors{Chen et al.}

\usepackage{color}

\graphicspath{{./}{figures/}}
\begin{document}
\title{FRBs Lensed by Point Masses I. Lens Mass Estimation for Doubly Imaged FRBs}
\email{Contact e-mail: xcchen@pmo.ac.cn (X. Chen);  ypshu@mpa-garching.mpg.de (Y. Shu)}
\author{Xuechun Chen}
\affil{Purple Mountain Observatory, Chinese Academy of Sciences, Nanjing, Jiangsu, 210023, China}
\affil{School of Astronomy and Space Science, University of Science and Technology of China, Hefei, Anhui, 230026, China}
\author{Yiping Shu}
\affil{Max-Planck-Institut f\"{u}r Astrophysik, Karl-Schwarzschild-Str. 1, 85748 Garching, Germany}
\affil{Ruhr University Bochum, Faculty of Physics and Astronomy, Astronomical Institute (AIRUB), German Centre for Cosmological Lensing, 44780 Bochum, Germany}
\author{Wenwen Zheng}
\affil{Purple Mountain Observatory, Chinese Academy of Sciences, Nanjing, Jiangsu, 210023, China}
\affil{School of Astronomy and Space Science, University of Science and Technology of China, Hefei, Anhui, 230026, China}
\author{Guoliang Li}
\affil{Purple Mountain Observatory, Chinese Academy of Sciences, Nanjing, Jiangsu, 210023, China}

\begin{abstract}
Fast radio bursts (FRBs) are bright radio transient events with durations on the order of milliseconds. The majority of FRB sources discovered so far have a single peak, with the exception of a few showing multiple-peaked profiles, the origin of which is unknown. In this work, we show that the strong lensing effect of a point mass or a point mass $+$ external shear on a single-peak FRB can produce double peaks (i.e. lensed images). In particular, the leading peak will always be more magnified and hence brighter than the trailing peak for a point-mass lens model, while the point-mass $+$ external shear lens model can produce a less magnified leading peak. We find that, for a point-mass lens model, the combination of lens mass $M$ and redshift $z_l$ in the form of $M(1+z_l)$ can be directly computed from two observables---the delayed time $\Delta t$ and the flux ratio of the leading peak to the trailing peak $R$. For a point-mass $+$ external shear lens model, upper and lower limits in $M(1+z_l)$ can also be obtained from $\Delta t$ and $R$ for a given external shear strength. In particular, tighter lens mass constraints can be achieved when the observed $R$ is larger. Lastly, we show the process of constraining lens mass using the observed values of $\Delta t$ and $R$ of two double-peaked FRB sources, i.e. FRB 121002 and FRB 130729, as references, although the double-peaked profiles are not necessarily caused by strong lensing.
\end{abstract}


\section{Introduction}
\label{sec:intro}
Fast radio bursts (FRBs) are very bright radio transient signals with durations of milliseconds. They have been detected from several hundred MHz to several GHz. Since the first discovery in 2007 \citep{Lorimer2007}, there have been more than 100 FRBs captured by Parkes, UTMOST, ASKAP, CHIME, FAST, etc.\footnote{http://www.frbcat.org} \citep[e.g.,][]{Petroff2016, Zhu2020}. The arrival times of an FRB in different frequencies are different. The difference is proportional to the dispersion measure (DM), which is defined as the integration of free electron column density along the line of sight. It is noted that the DMs of many FRBs are far higher than the expected contribution from the Milky Way \citep[e.g.,][]{Cordes2002,Petroff2014}, suggesting that they are from extragalactic sources. In addition, FRB 121102 \citep{Gajjar2018} has been precisely localized to a dwarf galaxy at $z=0.193$ \citep{Bassa2017,Chatterjee2017,Marcote2017,Tendulkar2017}, which further supports the extragalactic origin of FRBs.

The physical nature of FRBs has not been fully understood. Nevertheless, their characteristic properties such as being point-like, short duration, high event rate \citep[e.g.][]{Thornton2013,Champion2016} and extragalactic origin make FRBs a powerful probe for studying cosmology. For example, time-delay measurements in galaxy-scale, strong-lensed quasars have enabled independent, precise determinations of the Hubble constant \citep{Bonvin2017,Wong2020}. The durations of FRBs are much shorter than the typical time-delay scales in galaxy-scale strong lenses, which can lead to more precise time-delay measurements and therefore serve as another compelling cosmological probe \citep[e.g.,][]{Munoz2016,Li2017,Wagner2018,Liao2020,Sammons2020}.

The majority of currently known FRB sources show a single burst, while some are found to emit multiple bursts, i.e. repeating FRBs, that are separated by a wide range of time scales from seconds to years \citep{Spitler2016,CHIME2019,2019ApJ...885L..24C,Fonseca2020}. It suggests that at least some fraction of FRBs are not produced by cataclysmic events, although the actual physical mechanisms of repeating FRBs are still unclear. In fact, whether there is any truly nonrepeating FRB source is also open for discussions. On the one hand, it is possible that some single-burst FRB sources are also repeaters, but only one burst has been captured due to various factors because, for example, the broad range of repetition rates or the other bursts are simply too faint \citep{2018Natur.562..386S}. On the other hand, some studies seem to suggest that nonrepeating and repeating FRB sources have different physical properties. For example, the burst widths of repeating FRB sources are found to be generally wider than those of nonrepeating FRB sources \citep[e.g.,][]{2019ApJ...885L..24C}. Several repeating FRB sources are also found to exhibit similarly downward frequency drifts of the subbursts \citep{Gajjar2018,2019ApJ...876L..23H,2019ARA&A..57..417C,Petroff2019}. Nevertheless, those results are based on a small sample of repeating FRB sources. A clearer picture will emerge as the sample of FRB sources increases in the near future.

Strong gravitational lensing provides an additional way of producing multiple peaks from a single peak. The creative idea that compact objects such as massive compact halo objects (MACHOs) or primordial black holes (PBHs) can lens FRBs into double-peaked profiles was proposed by \citet{Munoz2016}. \citet{Liao2020} and \citet{Sammons2020} also suggested that the fraction of compact objects in dark matter can be constrained with multiple images of lensed FRBs. Under the weak-field approximation, some objects such as a Schwarzschild black hole (BH) can be treated as a point-mass lens. The point-mass lens is a classical model that can produce a double-peaked signal (i.e. two lensed images), with the leading peak always brighter (i.e. more magnified) than the trailing peak. Clearly, such a model cannot explain a brighter trailing peak. In this work, we show that a brighter trailing peak can be produced when an additional external shear is present. Moreover, we derive some basic properties of a point-mass lens model and a point-mass $+$ external shear lens model and discuss how to constrain the lens mass from observables including the time delay and flux ratio. For the purpose of illustration, we show two examples where time delays and flux ratios are taken as the observed time delays and flux ratios in the two known double-peaked FRB sources: FRB 121002 and FRB 130729. Their relevant parameters are shown in Table1. Nevertheless, we note that double peaks from those two sources are not necessarily due to gravitational lensing. 

The paper is organized as follows. Section~\ref{sec:PointMass} presents the basic properties and lens mass constraints for a point mass-lens model, and Section~\ref{sec:pointshear} presents the basic properties and lens mass constraints for a point-mass $+$ external shear lens model. We provide some discussions in Section~\ref{sec:discussion}. A summary is given in Section~\ref{sec:conclusion}. Throughout the paper, we assume a fiducial cosmology of $\Omega_m = 0.272, \Omega_{\lambda} = 0.728$, and $H_0=70.4$ km s$^{-1}$ Mpc$^{-1}$\citep{Komatsu2011}.

\section{Point mass lens model}
\label{sec:PointMass}

\subsection{Basic properties}

The theory of point-mass lens model is detailed in \citet{Schneider1986} and \citet{PJ1992}. For a point mass of $M$, its Einstein angle is defined as
\begin{eqnarray}
\label{eq:thetaE}
    \theta _{E}=\sqrt{\frac{4GM}{c^{2}}\, \frac{D_{ls}}{D_{l}D_{s}}}.
\end{eqnarray}
The lens potential of a point-mass lens model is
\begin{eqnarray}
    \psi (\theta )=\theta _{E}^{2}\ln(\theta ),
\end{eqnarray}
and the lens equation is
\begin{eqnarray}
    \beta =\theta -\theta _{E}^{2}\frac{\theta }{|\theta |^{2}},
\end{eqnarray}
where $\beta$ is the position in the source plane and $\theta$ is the position in the deflector plane, and $D_{l}$, $D_{s}$ and $D_{ls}$ are the angular diameter distances from the lens to the observer, from the source to the observer, and from the source to the lens respectively. The lens equation scaled by $\theta_{E}$, where $y=\frac{\beta}{\theta_{E}}$, $x=\frac{\theta}{\theta_{E}}$, becomes
\begin{eqnarray}
    y=x-\frac{1}{x},
\end{eqnarray}
where $y$ is a dimensionless quantity called the collision parameter. For a point-mass lens, two images are produced for a source located at angular position $\beta \neq 0$ and the two solutions of the lens equation: $x_{\pm }=\frac{1}{2}(y \pm \sqrt{y^{2}+4})$ on opposite sides of the lens with a separation angle of $\theta_{d}$. $"+"$ and $"-"$ denote the parity of the image. Note that in the case of a source lying exactly on the optical axis (i.e. $\beta=0$), a full ring instead of two images appears because of rotational symmetry. In general, the opening angle between the two lensed images is
\begin{eqnarray}
\label{eq:thetad}
    \theta _{d}=\sqrt{y^{2}+4}\theta _{E}.
\end{eqnarray}
The time-delay function $t$ at the two lensed images is
\begin{eqnarray}
\label{eq:time-delay-function-pm}
    t(x_{\pm })=\frac{4GM}{c^{3}}(1+z_{l})(\frac{2y^{2}+4\mp 2y\sqrt{y^{2}+4}}{8}-\ln\left | x_{\pm} \right |),
\end{eqnarray}
where $z_{l}$ is the redshift of the lens.
The time delay of image $x_{-}$ with respect to $x_{+}$ is, therefore,
\begin{eqnarray}
\label{eq:time-delay-pm}
\begin{aligned}
    \Delta t&=t(x_{-})-t(x_{+})\\&=\frac{4GM}{c^{3}}(1+z_{l})\left [ \frac{y\sqrt{y^{2}+4}}{2}+\ln(\frac{\sqrt{y^{2}+4}+y}{\sqrt{y^{2}+4}-y}) \right ].
\end{aligned}
\end{eqnarray}
For a point-mass lens model, $x_{+}$ is always the leading image.
Their magnifications are
\begin{eqnarray}
\label{eq:mu-pm}
    \mu _{\pm }=\pm \frac{1}{4} \left [ \frac{y}{\sqrt{y^{2}+4}}+\frac{\sqrt{y^{2}+4}}{y}\pm 2 \right ],
\end{eqnarray}
respectively. We can calculate the flux ratio of the two images and define $R$ as the flux ratio of the leading image to the trailing image, i.e. the leading-to-trailing flux ratio:
\begin{eqnarray}
\label{eq:flux-ratio-pm}
    R=\left | \frac{\mu _{+}}{\mu _{-}} \right |=\frac{y^{2}+2+y\sqrt{y^{2}+4}}{y^{2}+2-y\sqrt{y^{2}+4}}.
\end{eqnarray}

\begin{deluxetable}{cccccc}
\tablenum{1}
\tablecaption{Some parameters of FRB 121002 and FRB 130729.\label{tab:table1}}
\tablewidth{\textwidth}
\tablehead{
\colhead{ID} & \colhead{R} & \colhead{$\Delta t$(ms)} & \colhead{Redshift} &
\colhead{DM(pc\,$cm^{-3}$)} & \colhead{Telescope} \\
}
\startdata
121002 & 0.91 & 2.4 & 1.3 & 1629 & parkes \\
130729 & 1.95 & 11 & 0.69 & 861 & parkes \\
\enddata
\end{deluxetable}

\begin{figure}
\centerline{\scalebox{1.0}
{\includegraphics[width=0.48\textwidth]{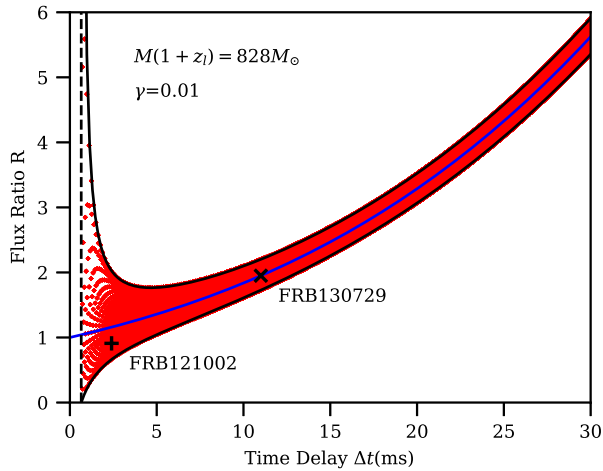}}}
\caption{\label{fig:R-Deltat} Distributions of the leading-to-trailing flux ratio $R$ and time delay $\Delta t$ for a point-mass and a point-mass $+$ external shear lens model. The blue solid line corresponds to the $R$-$\Delta t$ relation for a point-mass lens model with $M(1+z_{l})=828M_{\odot}$. The red crosses show possible pairs of ($R$, $\Delta t$) obtained from a forward ray-tracing simulation for a point-mass $+$ external shear lens model with $M(1+z_{l})=828M_{\odot}$ and $\gamma=0.01$. The black cross and black plus mark the observed time delays and leading-to-trailing flux ratios of FRB 130729 and FRB 121002 respectively. The vertical black dashed line and two black solid curves represent the left, upper, and lower boundaries of all possible ($\Delta t$, $R$) pairs for a point-mass $+$ external shear lens model with $M(1+z_{l})=828M_{\odot}$ and $\gamma=0.01$. Please see Section~\ref{sec:pointshear} for a detailed explanation. }
\end{figure}

In the point-mass lens model, the image arriving earlier is called the Fermat minimum image while the other is called the Fermat saddle image. From Equation (\ref{eq:time-delay-pm}) and Equation (\ref{eq:flux-ratio-pm}), one can see that the Fermat minimum image always has a higher magnification. That means $R$ is always $\geqslant 1$. It is also reflected by the blue solid line in Figure \ref{fig:R-Deltat}.

If we combine Equation (\ref{eq:time-delay-pm}) and Equation (\ref{eq:flux-ratio-pm}) to eliminate the collision parameter $y$, the mass of the point lens can be expressed as
\begin{eqnarray}
\label{eq:mass-z}
\begin{aligned}
    M(1+z_{l})&=\frac{c^{3}\Delta t}{2G}\frac{\sqrt{R}}{R-1+\sqrt{R} \ln R}\\
    &\approx 101.5\frac{\Delta t}{1\text{ms}}\frac{\sqrt{R}}{R-1+\sqrt{R}\ln R}M_{\odot}.
\end{aligned}
\end{eqnarray}
In the scenario of an FRB lensed by a point-mass object, the mass of the lens can be determined from the observables: $\Delta t, R, $ and $z_{l}$. We note that the mass estimation above is not valid for $y=0$, in which case the lensed image is a circularly symmetric Einstein ring with no relative time delay.

\begin{figure}
\centerline{\scalebox{1.0}
{\includegraphics[width=0.48\textwidth]{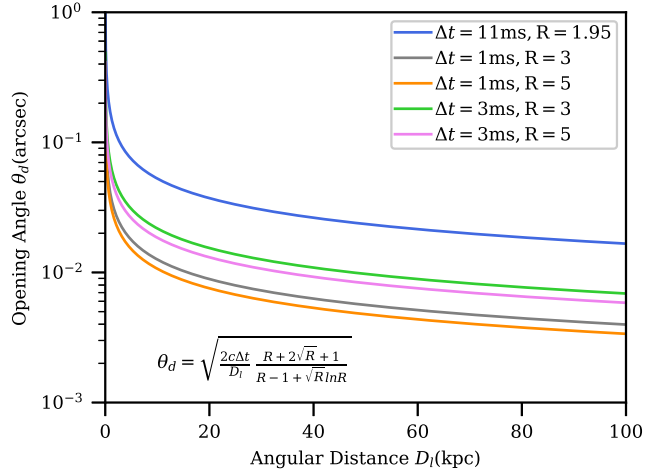}}}
\caption{\label{fig:thetad-dl}The opening angle $\theta_{d}$ as a function of the lens distance $D_{l}$ for various flux ratio and time-delay combinations.}
\end{figure}

In addition, by combining Equation (\ref{eq:thetad}) and Equation (\ref{eq:flux-ratio-pm}), we can rewrite the opening angle between the two lensed images as
\begin{eqnarray}
    \theta _{d}=\frac{\sqrt{R}+1}{R^{1/4}}\theta _{E}.
\end{eqnarray}
Substituting Einstein angle and lens mass by Equation (\ref{eq:thetaE}) and Equation (\ref{eq:mass-z}), we obtain
\begin{eqnarray}
\begin{aligned}
    \theta _{d}&=\sqrt{\frac{2c\Delta t D_{ls}}{(1+z_{l})D_{l}D_{s}}\frac{R+1+2\sqrt{R}}{R-1+\sqrt{R}\ln R}}\\
    &\approx 0.91\sqrt{\frac{\Delta t}{1\text{ms}}}\sqrt{\frac{D_{ls}\times 1Mpc}{(1+z_{l})D_{l}D_{s}}}\sqrt{\frac{R+1+2\sqrt{R}}{R-1+\sqrt{R}\ln R}} \text{  mas}.
\end{aligned}
\end{eqnarray}
The above equation implies that if the opening angle of the two lensed images can be determined, one can estimate the distance or redshift of the lens. Figure \ref{fig:thetad-dl} shows the opening angle $\theta_{d}$ as a function of the lens distance for various flux ratio and time-delay combinations. Here we actually show an approximate relation assuming that the FRB source is at a much higher redshift than the lens and hence $D_{ls} \approx D_{s}$. The blue curve shows the opening angle function for a lensed FRB with the same time delay and flux ratio as FRB 130729. The typical opening angle of $0.01^{''}-0.1^{''}$ is so small that most currently radio telescopes have a hard time resolving the lensed images. Nevertheless, a lower limit of the lens redshift $z_{l}$ can still be derived from the resolution of the observations if the two lensed images are unresolved.

\subsection{Lens Mass Estimation}
\label{sec:BHMass}
For a point-mass lens, one can accurately determine the mass of the lens from the separation of lensed images given the lens and source redshifts. However, these three quantities are all hard to measure because of the small image separation, the lack of electromagnetic counterparts if the lens is an intermediate-mass black hole (IMBH), and the uncertainty of the FRB redshift inferred from the DM. Alternatively, as we show in Equation (\ref{eq:mass-z}), the lens mass can be computed from two other easily obtained observables: the time delay $\Delta t$ and leading-to-trailing flux ratio of the two lensed peaks $R$.

For example, FRB 130729 \citep{Champion2016} is a double-peaked FRB source at redshift $z_{s}=0.69$ with an observed leading-to-trailing flux ratio of 1.95 and a time delay of 11 ms. Imagining a single-peak FRB being lensed by a point mass into two images with the same leading-to-trailing flux ratio and time delay as FRB 130729, the mass of the lens can be directly computed as
\begin{eqnarray}
    M (z_l) =\frac{828}{1+z_{l}}M_{\odot }.
\end{eqnarray}
The blue solid line in Figure \ref{fig:R-Deltat} shows the relation between $R$ and $\Delta t$ in Equation (\ref{eq:mass-z}) with $M(1+z_{l})=828M_{\odot}$. The black cross marks the observed $R$ and $\Delta t$ of FRB 130729. The mass estimation is not sensitive to the redshift of the lens as it scales as $1/(1+z_{l})$. If the lens resides within the Milky Way (i.e., $z_{l}\approx 0$), its mass would be about $828M_{\odot}$. A lower limit of the lens mass can be obtained if the redshift of the FRB source can be determined, for instance, from the dispersion measure or optical counterpart. Adopting source redshift of 0.69 (the redshift of FRB 130729) as the upper limit of $z_l$, the lower limit of the intervening lens mass is about $490M_{\odot}$. Such mass scales can be achieved for IMBHs.

The uncertainty of the lens mass is contributed from three quantities:
\begin{enumerate}
    \item For time delay $\Delta t$, it has $\frac{\delta M}{M}=\frac{\delta \Delta t}{\Delta t}$. This uncertainty could be tiny when the time resolution is much less than the time delay between the two images.
    \item For flux ratio $R$, it has $\frac{\delta M}{M}=\frac{(1+\sqrt{R})^{2}}{2(R-1+\sqrt{R}\ln R)}\frac{\delta R}{R}$. The first term on the right hand is about 1.5 for $R$=1.95 and quickly drops to below 1 when $R>2.85$. Therefore the relative uncertainty in mass is comparable with $\frac{\delta R}{R}$ and decreases with increased $R$.
    \item For redshift of the lens $z_{l}$, two scenarios should be considered. When lens redshift can be measured, it has $\frac{\delta M}{M}=\frac{z_{l}}{1+z_{l}}\frac{\delta z_{l}}{z_{l}}$. This contribution is expected to be small because $\delta z_l$ is generally very small when $z_l$ can be measured. On the other hand, when the lens redshift is unknown because either it cannot be directly measured or the lensed images are unresolved, one can choose to set the lens redshift to $z_l = \frac{z_s (2z_s+3)}{3(z_s+2)}$. This redshift choice minimizes the integral of $[M(z_l)/M(z_{\rm true})-1]^2$ over $z_{\rm true}=0$--$z_s$ where $z_{\rm true}$ is the true lens redshift, and the largest possible uncertainty in the lens mass is reached when $z_{\rm true}$=0. For example, if the source redshift is 0.69 (i.e. the redshift of FRB 130729), one can set $z_l$ to 0.374, which leads to a lens mass of 603 $M_\odot$. The largest possible uncertainty is then 27.2\%, which will be the main source of uncertainty in the lens mass.
\end{enumerate}

\section{point-mass + external shear model}
\label{sec:pointshear}

As demonstrated above, for a point-mass lens model, the lensed image that arrives earlier always appears brighter than the one that arrives later. Although plasma lensing combined with gravitational lensing could reverse this trend, in this section, we show that by adding an external shear component, which is usually associated with potential perturbations, it is possible to produce a gravitational-lensing-induced double-peaked profile with $R < 1$.

\subsection{Basic properties}

The theory of point-mass $+$ external shear lens model has been studied in detail in \citet{Chang1979,Chang1984} and \citet{AnEvans2006} with the following effective lensing potential:
\begin{eqnarray}
\label{eq:psi-pmg}
    \psi(\theta)=\theta_{E}^{2}\ln(\theta)-\frac{\gamma_{1}}{2}(\theta_{1}^{2}-\theta_{2}^{2})+\gamma_{2}\theta_{1}\theta_{2},
\end{eqnarray}
where $\theta_{E}$ is defined by Equation (\ref{eq:thetaE}) and the external shear strength is defined as $\gamma=\sqrt{\gamma_{1}^{2}+\gamma_{2}^{2}}$. By rotating the coordinates, Equation (\ref{eq:psi-pmg}) can be rewritten as
\begin{eqnarray}
    \psi (\theta )=\theta_{E}^{2}\ln(\theta)-\frac{\gamma}{2}(\theta_{1}^2-\theta_{2}^2).
\end{eqnarray}
The lens equation is
\begin{eqnarray}
\begin{aligned}
    &\beta_{1}=\theta_{1}+\gamma \theta_{1}-\theta_{E}^{2}\frac{\theta_{1}}{|\theta|^{2}},\\
    &\beta_{2}=\theta_{2}-\gamma \theta_{2}-\theta_{E}^{2}\frac{\theta_{2}}{|\theta|^{2}}.
\end{aligned}
\end{eqnarray}
Scaling the angular coordinates with $\theta_{E}$: $y=\frac{\beta}{\theta_{E}}$, $x=\frac{\theta}{\theta_{E}}$, the lens equation reads in dimensionless form:
\begin{eqnarray}
\begin{aligned}
    &y_{1}=(1+\gamma)x_{1}-\frac{x_{1}}{x_{1}^{2}+x_{2}^{2}},\\
    &y_{2}=(1-\gamma)x_{2}-\frac{x_{2}}{x_{1}^{2}+x_{2}^{2}}.
\end{aligned}
\end{eqnarray}
The lens equation has possibly four solutions at most and is not easy to solve directly. Here we show the results along the axis (i.e., $y_{1}=0$ or $y_{2}=0$) for we can always get to this through coordinate rotation.
For $y_{2}=0$, we get
\begin{eqnarray}
\label{eq:along-y1-solution1}
\begin{aligned}
    &x_{1}=\frac{y_{1}\pm \sqrt{y_{1}^{2}+4(1+\gamma)}}{2(1+\gamma)},\\
    &x_{2}=0;
\end{aligned}
\end{eqnarray}
and
\begin{eqnarray}
\label{eq:along-y1-solution2}
\begin{aligned}
    &x_{1}=\frac{y_{1}}{2\gamma},\\
    &x_{2}=\pm \sqrt{\frac{1}{1-\gamma}-\frac{y_{1}^{2}}{4\gamma^{2}}}.
\end{aligned}
\end{eqnarray}
For $y_{1}=0$, we get
\begin{eqnarray}
\label{eq:along-y2-solution1}
\begin{aligned}
    &x_{1}=0,\\
    &x_{2}=\frac{y_{2}\pm \sqrt{y_{2}^{2}+4(1-\gamma)}}{2(1-\gamma)};
\end{aligned}
\end{eqnarray}
and
\begin{eqnarray}
\label{eq:along-y2-solution2}
\begin{aligned}
    &x_{1}=\pm \sqrt{\frac{1}{1+\gamma}-\frac{y_{2}^{2}}{4\gamma^{2}}},\\
    &x_{2}=-\frac{y_{2}}{2\gamma}.
\end{aligned}
\end{eqnarray}
It should be noted that the two roots of Equations (\ref{eq:along-y1-solution2}) and (\ref{eq:along-y2-solution2}) appear only when $|y_{1}|\leqslant \frac{2\gamma}{\sqrt{1-\gamma}}$, $|y_{2}|\leqslant \frac{2\gamma}{\sqrt{1+\gamma}}$, respectively.

We can calculate the magnification of each image with the Jacobian matrix \emph{A}:
\begin{eqnarray}
\begin{aligned}
    &\mu(\theta)=\frac{1}{det\emph{A}(\theta)},\\
    &\emph{A}(\theta)=\frac{\partial \beta}{\partial \theta}.
\end{aligned}
\end{eqnarray}
The magnification $\mu$ follows
\begin{eqnarray}
\label{eq:mu-pmg}
    \mu^{-1}=1-\gamma^{2}-\frac{1}{|x|^{4}}-2\gamma\frac{x_{1}^{2}-x_{2}^{2}}{|x|^{4}}.
\end{eqnarray}
The caustic points on the $y_{1}$-axis are $y_{1}=\pm \frac{2\gamma}{\sqrt{1-\gamma}}$ and the corresponding critical points on the $x_{1}$-axis are $x_{1}=\pm \frac{1}{\sqrt{1-\gamma}}$. Here we just consider the case when $\gamma\ll 1$. This means that the size of caustic is far less than $\theta_{E}$, and the cross section of four-image configurations is negligible. So, we only focus on two-image configurations in the following analysis.

Using Equations (\ref{eq:along-y1-solution1}) and (\ref{eq:mu-pmg}), we obtain the time delay
\begin{eqnarray}
\label{eq:time-delay-pmg}
    \Delta t=\frac{4GM}{c^{3}}(1+z_{l})\left [ \frac{y_{1}s}{2(1+\gamma)}+\ln\left ( \frac{s+y_{1}}{s-y_{1}}\right )\right ],
\end{eqnarray}
and the flux ratio
\begin{eqnarray}
\label{eq:flux-ratio-pmg}
    R=|\frac{\mu_{+}}{\mu_{-}}|=\frac{[y_{1}^{2}+2+y_{1}s][y_{1}s+4\gamma]+8\gamma^{2}-2\gamma y_{1}^{2}}{[y_{1}^{2}+2-y_{1}s][y_{1}s-4\gamma]-8\gamma^{2}+2\gamma y_{1}^{2}},
\end{eqnarray}
where $s=\sqrt{y_{1}^{2}+4(1+\gamma)}$. The corresponding formulas for $y_{1}=0$ can be obtained just by simply replacing $y_{1}$ with $y_{2}$ and changing $\gamma$ to $-\gamma$, i.e.,
\begin{eqnarray}
\label{eq:time-delay-pmg-y2}
    \Delta t=\frac{4GM}{c^{3}}(1+z_{l})\left [ \frac{y_{2}s^{\prime}}{2(1-\gamma)}+\ln\left ( \frac{s^{\prime}+y_{2}}{s^{\prime}-y_{2}}\right )\right ],
\end{eqnarray}
and the flux ratio
\begin{eqnarray}
\label{eq:flux-ratio-pmg-y2}
    R=\frac{[y_{2}^{2}+2+y_{2}s^{\prime}][y_{2}s^{\prime}-4\gamma]+8\gamma^{2}+2\gamma y_{2}^{2}}{[y_{2}^{2}+2-y_{2}s^{\prime}][y_{2}s^{\prime}+4\gamma]-8\gamma^{2}-2\gamma y_{2}^{2}},
\end{eqnarray}
where $s^{\prime}=\sqrt{y_{2}^{2}+4(1-\gamma)}$.

\subsection{Lens Mass Constraints}

For a point-mass $+$ external shear lens model, the lens mass generally cannot be unambiguously determined from the flux ratio and time delay because the number of unknowns is more than the number of equations. Nevertheless, it is still possible to obtain some constraints on the lens mass. In particular, we focus on doubly imaged configurations as we are interested in FRBs with two peaks. We perform a forward ray-tracing simulation for a point-mass $+$ external shear model with $M(1+z_l)=828 M_\odot$ and $\gamma=0.01$, and plot $R$-$\Delta t$ pairs for all the simulated doubly-imaged cases in Figure~\ref{fig:R-Deltat} as red crosses. We find that all possible $R$-$\Delta t$ pairs between the two lensed images are actually bracketed by three black curves in Figure~\ref{fig:R-Deltat}, which we now explain in detail.

The dashed vertical line corresponds to the lower limit in $\Delta t$ for a fixed $M(1+z_l)$ and $\gamma$, when the source is located at the tips of the inner caustics, i.e. $y_1=0, y_2=\pm \frac{2\gamma}{\sqrt{1+\gamma}}$ or $y_1=\pm \frac{2\gamma}{\sqrt{1-\gamma}}, y_2=0$. This lower limit can be written as
\begin{align}
\label{eq:M_min_left}
\Delta t_{\rm min} &= \frac{4 G M}{c^3} (1+z_l) (\frac{2 \gamma}{1-\gamma^2} + \ln \frac{1+\gamma}{1-\gamma}) \\
& \approx \frac{M(1+z_l)}{1268.5 M_{\odot}} \frac{\gamma}{0.01} \text{ ms } (\text{when } \gamma \ll 1).
\end{align}
The corresponding leading-to-trailing flux ratio $R$ is 0 or $+\infty$. We use a dashed line because the lower limit in $\Delta t$ is only reached when $R$ is 0 or $+\infty$. The top of the two solid black curves corresponds to the $R$-$\Delta t$ relation along the $y_1$-axis (i.e. $y_2=0$). This $y_2=0$ curve approaches $+ \infty$ on both ends and reaches its minimum at $y_{1}=\sqrt{\frac{2\gamma\left ( 1+2\gamma\right )}{1-\gamma}}$ with
\begin{eqnarray}
R_{\rm min}=\frac{\left [ 4\gamma^{2}+2+\Delta \right ]\left [ \Delta+4\gamma-4\gamma^{2}\right ]+16\gamma^{4}-20\gamma^{3}+4\gamma^{2}}{\left [ 4\gamma^{2}+2-\Delta \right ]\left [ \Delta-4\gamma+4\gamma^{2}\right ]-16\gamma^{4}+20\gamma^{3}-4\gamma^{2}},
\end{eqnarray}
where $\Delta=\sqrt{8\gamma^{3}+20\gamma^{2}+8\gamma}$. It is clear that this minimum $R$ value increases with $\gamma$ and is always larger than 1 when $\gamma > 0$. The bottom of the two solid black curves corresponds to the $R$-$\Delta t$ relation along the $y_2$-axis (i.e. $y_1=0$), which increases monotonically from $R=0$ to $R=+ \infty$. We find that these two solid black curves gradually converge to a narrow, rising band as $R$ increases. The rising speed of the band decreases with increasing lens mass $M(1+z_l)$. In the limit of $\gamma = 0$, these three curves will merge into one curve, which is the solid blue curve determined by Equation~(\ref{eq:mass-z}).

The exact shape of the ``permitted area'' in the $R$-$\Delta t$ space will change with both $M(1+z_l)$ and $\gamma$. At a fixed $\gamma$, the permitted area will move entirely toward the positive $\Delta t$ direction and the band enclosed by the $y_1=0$ and $y_2=0$ curves will become flatter as $M(1+z_l)$ increases. At a fixed $M(1+z_l)$, the permitted area will become fatter along the $R$ dimension as $\gamma$ increases. As a result, a given pair of observed $R_{\rm obs}$ and $\Delta t_{\rm obs}$ can only be found in the permitted areas of certain mass ranges for a certain $\gamma$, which leads to constraints on the lens mass. We now discuss those constraints one by one.

First of all, the left boundary of the permitted area translates into a pseudo upper limit in $M (1+z_l)$ that is given by Equation~(\ref{eq:M_min_left}), or $1268.5 (\frac{\Delta t_{\rm obs}}{1 \text{ms}}) (\frac{0.01}{\gamma}) M_\odot$ when $\gamma \ll 1$. We refer to such a limit as a pseudo limit because it also depends on $\gamma$. Obviously, this pseudo upper limit, denoted as $M_{\rm limit}^{\rm left}$, always exists, regardless of the values of $R_{\rm obs}$ and $\Delta t_{\rm obs}$.

Second, for a given $\gamma$, ($\Delta t_{\rm obs}$, $R_{\rm obs}$) can fall below the lower boundary of the permitted area when $M (1+z_l)$ is below a certain value. It implies that a pseudo lower limit in $M (1+z_l)$, denoted as $M_{\rm limit}^{\rm lower}$, always exists. The exact value of $M_{\rm limit}^{\rm lower}$ is related to $\gamma$ by Equations~(\ref{eq:time-delay-pmg-y2}) and (\ref{eq:flux-ratio-pmg-y2}).

Lastly, the upper boundary of the permitted area may impose additional constraints in $M(1+z_l)$. Because of the U shape of the upper boundary, we need to consider two scenarios. When $\gamma$ becomes so large that $R_{\rm min} \geqslant  R_{\rm obs}$, the observed ($\Delta t_{\rm obs}$, $R_{\rm obs}$) can always remain below the upper boundary, and no additional constraints on $M(1+z_l)$ is available. In particular, we note that $R_{\rm obs} < 1$ always belongs to this scenario because $R_{\rm min}$ is always no smaller than 1. When instead $R_{\rm min} < R_{\rm obs}$, for a given $\gamma$, ($\Delta t_{\rm obs}$, $R_{\rm obs}$) will first move out of and then move into the permitted area (in a relative sense) as $M(1+z_l)$ increases from $M_{\rm limit}^{\rm lower}$ to $M_{\rm limit}^{\rm left}$. As a result, the permitted mass range of $[M_{\rm limit}^{\rm lower}, M_{\rm limit}^{\rm left}]$ will be divided into two smaller ranges. The first intermediate-mass scale, denoted as $M_{\rm limit}^{\rm rising}$, corresponds to the situation where ($\Delta t_{\rm obs}$, $R_{\rm obs}$) moves onto the rising part of the upper boundary of the permitted area, and the other intermediate-mass scale, denoted as $M_{\rm limit}^{\rm falling}$, corresponds to the situation where ($\Delta t_{\rm obs}$, $R_{\rm obs}$) moves onto the falling part of the upper boundary of the permitted area. $M_{\rm limit}^{\rm falling}$ and $M_{\rm limit}^{\rm rising}$ can be obtained from Equations~(\ref{eq:time-delay-pmg}) and (\ref{eq:flux-ratio-pmg}). Consequently, the pseudo limits in $M(1+z_l)$ becomes to $[M_{\rm limit}^{\rm lower}, M_{\rm limit}^{\rm rising}]$ and $[M_{\rm limit}^{\rm falling}, M_{\rm limit}^{\rm left}]$.

\begin{figure*}
\centering
\includegraphics[width=0.98\textwidth]{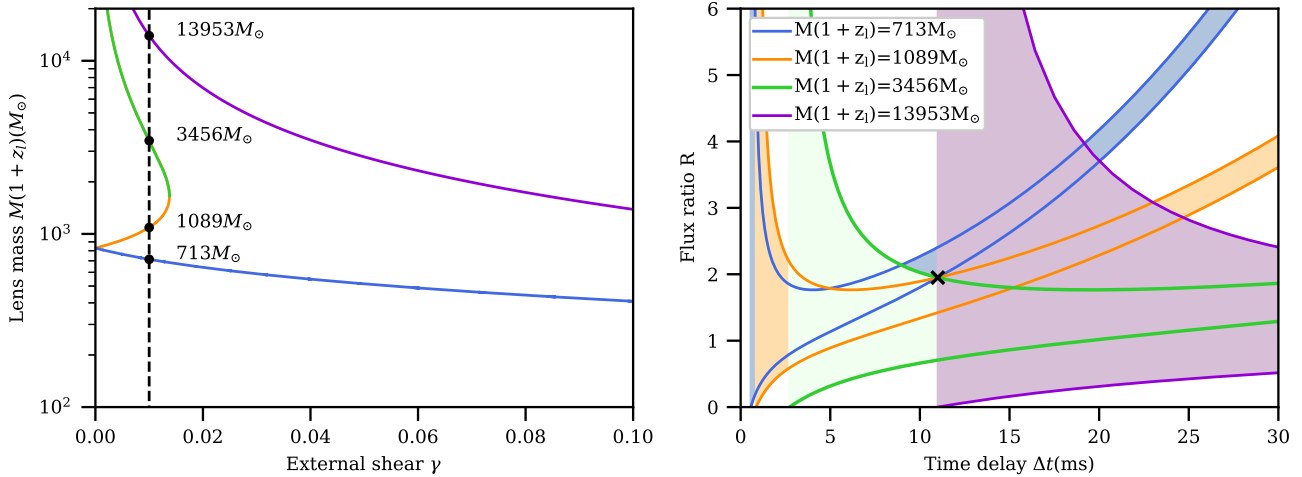}
\caption{\label{fig:mass-g} \rm Left: $M(1+z_l)$-$\gamma$ relations that corresponds to situations when ($R$=1.95, $\Delta t=11$ ms) falls on the left (violet line), upper (orange/green line), and lower (blue line) boundaries of the permitted area in the $R$-$\Delta t$ space. Intersections of a vertical $\gamma=0.01$ line and the three $M(1+z_{l})$-$\gamma$ relations indicate that two possible mass ranges exist in this case. \rm Right: Permitted areas in the $R$-$\Delta t$ space that correspond to the four intersecting masses seen in the left panel and $\gamma=0.01$.}
\end{figure*}

\begin{figure*}
\centerline{\scalebox{1.0}
{\includegraphics[width=0.98\textwidth]{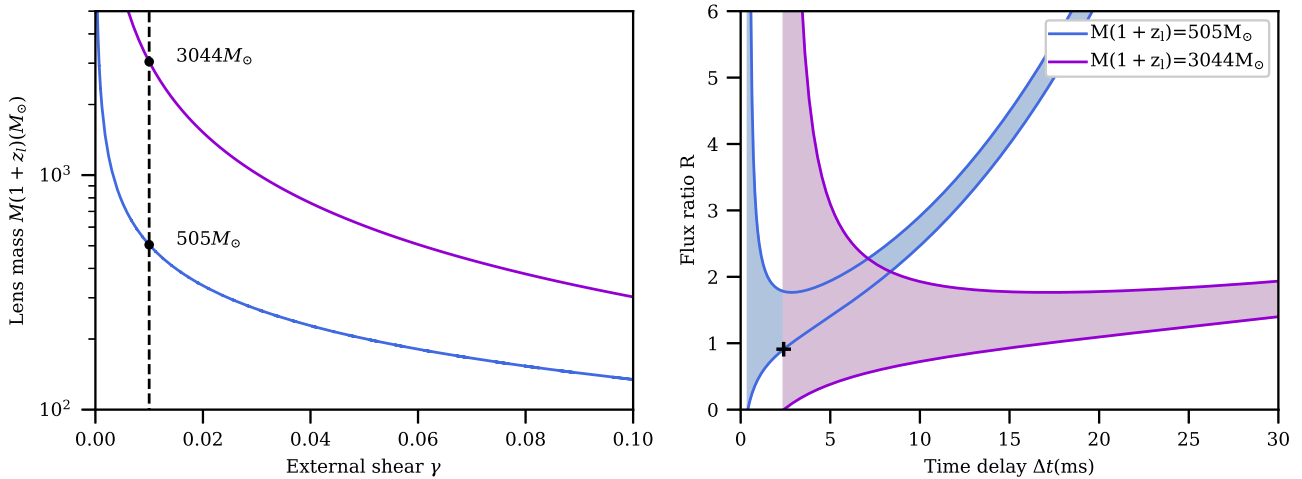}}}
\caption{\label{fig:R-Deltat-2} The same as Figure~\ref{fig:mass-g} but for $R$=0.91, $\Delta t=2.4$ ms. In this case, only two $M(1+z_{l})$-$\gamma$ relations are seen because the other one defined by the upper boundary does not exist.}
\end{figure*}

We now use the observed $R$ and $\Delta t$ values of the two double-peaked FRB sources to demonstrate how to obtain these pseudo mass constraints. For a doubly lensed FRB with $R$ and $\Delta t$ equal to those of FRB 130729, i.e. $R=1.95$, $\Delta t = 11$ ms, the three curves in the left panel of Figure~\ref{fig:mass-g} show the $M(1+z_l)$-$\gamma$ relations derived from Equations~(\ref{eq:time-delay-pmg}) and (\ref{eq:flux-ratio-pmg}) (orange/green line), Equations~(\ref{eq:time-delay-pmg-y2}) and (\ref{eq:flux-ratio-pmg-y2}) (blue line), and Equation~(\ref{eq:M_min_left}) (violet line), respectively. We find that the $M(1+z_l)$-$\gamma$ relation corresponding to the left boundary (violet line) monotonically decreases with $\gamma$. The $M(1+z_l)$-$\gamma$ relation corresponding to the lower boundary (blue line) monotonically decreases and gradually plateaus at large $\gamma$ values. The $M(1+z_l)$-$\gamma$ relation corresponding to the upper boundary first increases with $\gamma$ (orange line) and then turns over toward$+\infty$ as $\gamma$ approaches zero (green line). The turning point corresponds to the $\gamma$ value that leads to $R_{\rm min}=1.95$. The blue and orange lines intersect at $\gamma=0$. Constraints on $M(1+z_l)$ are given by intersections of a constant $\gamma$ line and the three $M(1+z_l)$-$\gamma$ relations. For example, if $\gamma=0$, the only intersection is found at $M(1+z_l)=828 M_\odot$, which falls back to the solution of the point-mass model as expected. For a moderate external shear strength of $\gamma=0.01$ (dashed line in the left panel of Figure~\ref{fig:mass-g}), the corresponding $R_{\rm min} \approx 1.76$ is smaller than 1.95. Hence, there exist two possible mass ranges that are defined by the four intersections, which correspond to $M_{\rm limit}^{\rm lower} (713 M_\odot)$, $M_{\rm limit}^{\rm rising} (1089 M_\odot)$, $M_{\rm limit}^{\rm falling} (3456 M_\odot)$, and $M_{\rm limit}^{\rm left} (13953 M_\odot)$ from bottom to top. If $\gamma$ is larger than the critical value at the turning point (i.e., $\gamma \approx 0.014$), the possible mass range will expand to $[713 M_\odot, 13953 M_\odot]$. Adopting the redshift of FRB 130729 ($z_s=0.69$) as the upper limit of $z_l$, the lower mass limit is therefore 422$M_\odot$ when $\gamma=0.01$, comparable to the lower mass limit inferred previously from a point-mass lens model. The permitted areas defined by the four mass bounds at $\gamma=0.01$ are plotted in the right panel of Figure~\ref{fig:mass-g} together with the reference point of (11 ms, 1.95). It shows how the reference point moves (in a relative sense) from the lower boundary of the permitted area to the rising part of the upper boundary and from the falling part of the upper boundary to the left boundary.

\begin{figure*}
\centerline{\scalebox{1.0}
{\includegraphics[width=0.98\textwidth]{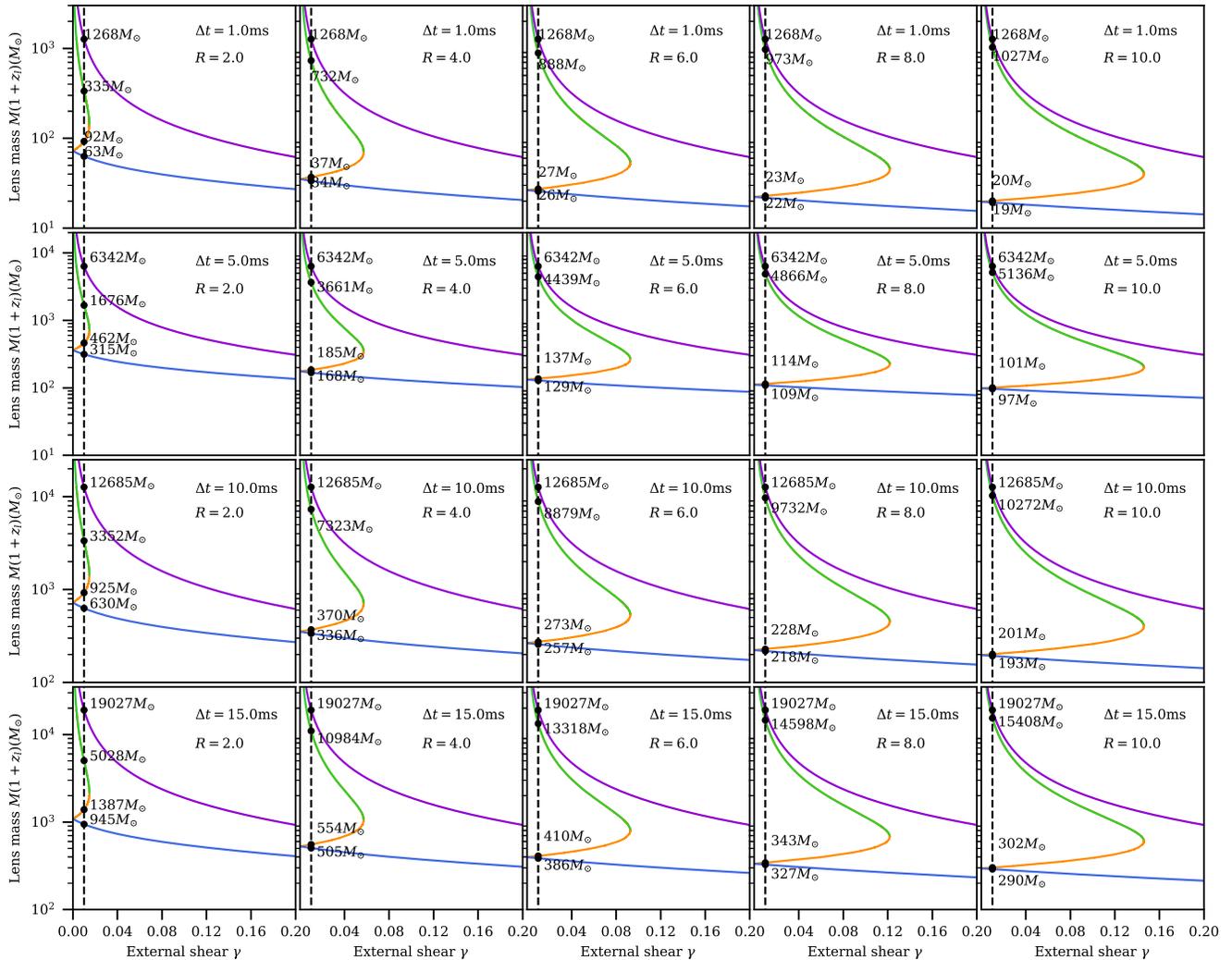}}}
\caption{\label{fig:trbehavior}The distribution of lens mass $M(1+z_{l})$ as a function of the external shear $\gamma$ for different pairs of flux ratio $R$ and time delay $\Delta t$. The meaning of the colored lines is consistent with that in Figure \ref{fig:mass-g} and Figure \ref{fig:R-Deltat-2}. The four mass intersections with a constant external shear $\gamma$ (dashed line) correspond to $M_{\rm limit}^{\rm lower}$, $M_{\rm limit}^{\rm rising}$, $M_{\rm limit}^{\rm falling}$, and $M_{\rm limit}^{\rm left}$ from bottom to top.}
\end{figure*}

For a doubly lensed FRB with $R$ and $\Delta t$ equal to those of FRB 121002, i.e. $R=0.91$, $\Delta t = 2.4$ms, the possible mass range is given by $[M_{\rm limit}^{\rm lower}, M_{\rm limit}^{\rm left}]$ because 0.91 will always be smaller than $R_{\rm min}$ for any $\gamma$. Therefore, only two $M(1+z_l)$-$\gamma$ relations defined from the lower and left boundaries are seen in the left panel of Figure~\ref{fig:R-Deltat-2}. For a moderate shear of $\gamma=0.01$, the corresponding limits in $M(1+z_l)$ are $[505 M_\odot, 3044 M_\odot]$. Adopting the redshift of FRB 121002 ($z_s=1.3$) as the upper limit of $z_l$, the lower mass limit is therefore 220$M_\odot$ when $\gamma=0.01$. We note that the lower mass limit tends to plateau as $\gamma$ increases. For a substantial amount of external shear ($\gamma=0.1$), the lower mass limit is about $M(1+z_l) = 134 M_\odot$ in this case. Therefore, if lensing-induced double peaks have flux ratio and time delay similar to those observed in FRB 121002, the intervening lens mass is most likely an IMBH or a MACHO that is more massive than $\sim 100M_{\odot}$.

Moreover, we find that tighter constraints can be obtained when $R_{\rm obs}$ is large because the permitted area becomes very narrow toward the large $R_{\rm obs}$ direction. In Figure~\ref{fig:trbehavior}, we show the $M(1+z_l)$-$\gamma$ relations for a grid of ($R_{\rm obs}$, $\Delta t_{\rm obs}$) pairs. It is clear that varying $\Delta t_{\rm obs}$ does not affect the shapes of the three $M(1+z_l)$-$\gamma$ relations but rather introduces a constant shift in the $\log (M(1+z_l))$ direction. However, the lens mass constraints becomes progressively tighter as $R_{\rm obs}$ increases.


\section{Discussion}
\label{sec:discussion}
Given the existence of repeating FRB sources, extra features apart from the band-integrated temporal profile need to be examined in order to determine whether the detected multiple peaks are due to gravitational lensing, chromatic plasma lensing, differential dispersion delays, or the intrinsic repetitions of FRB sources. First of all, lensing-induced multiple peaks will be spatially separated, while repeating FRBs are supposed to come from the same location. For example, \citet{Marcote2017} obtained milliarcsecond-resolution observations for the repeating FRB source FRB 121102 and found no significant positional offset between the detected individual bursts. Nevertheless, for the point-mass lens case discussed here, the required angular resolution is on the order of 0.01$^{\prime \prime}$--0.1$^{\prime \prime}$, which is hard to achieve in current discovery observations. Another useful feature is that gravitational lensing in general is achromatic. Therefore, one can examine the dynamic spectrum, i.e. the intensity as a function of both time and frequency. For lensing-induced multiple peaks, their temporal profiles should look the same (up to a constant normalization) in all frequencies. It does not need to be the case for multiple peaks in repeating FRB sources. For example, studies have found that some of the bursts in the repeating FRB source FRB 121102 show distinct components or cover finite band widths \citep[e.g.,][]{Gajjar2018,2019ApJ...876L..23H}. In addition, the polarization of the two peaks induced by a point-mass lens should be roughly the same if not identical because the spatial separation between the two peaks is so small that any difference in the effects on the polarization is supposed to be very subtle.

Plasma along the line of sight can modulate the spectral and temporal profiles of FRBs \citep[e.g.,][]{2019ARA&A..57..417C,Petroff2019,2021arXiv210104907X}. Dispersion will occur because the velocity of a radio signal in plasma is frequency-dependent (inversely proportional to the square of frequency), which is seen in the dynamic spectra of essentially all known FRB sources. The radio signal can be further scattered by particles in the plasma, which induces an asymmetric, frequency-dependent broadening to the burst profile \citep[e.g.,][]{Lorimer2007, 2019ARA&A..57..417C,Petroff2019}. Moreover, plasma inhomogeneity can act as a lens and introduce time delays and magnifications \citep{1998ApJ...496..253C,2012MNRAS.421L.132P,2017ApJ...842...35C,2020ApJ...889..158E}. A detailed comparison between gravitational lensing and plasma lensing has been given by \citet{Wagner&Er2020}. A key difference is that the plasma lensing effect is frequency dependent. In fact, plasma lensing has been suggested to be able to qualitatively explain many of the complex time--frequency structures and polarization observed in FRB 121102 \citep{2017ApJ...842...35C,2019ApJ...876L..23H}.

For a point-mass lens, the time delay $\Delta t$ between the two lensed FRB images will introduce a relative phase shift of $\Delta \phi = \omega \Delta t$, where $\omega = 2 \pi f$ is the frequency in the observer's frame. When the time resolution of the observation is better than the time delay $\Delta t$, the two lensed FRB images will not interfere as they will be separately resolved. For a time resolution of 10$\mu$s, it corresponds to a lens mass of the order of $0.1 M_{\odot}$ and above. When the lens mass is smaller than the characteristic mass scale given by the time resolution (according to Equation~\ref{eq:mass-z}), the two lensed FRB images will undergo constructive and destructive interference \citep[e.g.,][]{Zheng2014, Katz2020}. As a result, the total magnification (or intensity) of the two lensed images will oscillate with a frequency of $\Delta t^{-1}$, which corresponds to 1 MHz for a time delay of 1$\mu$s. This is well within the reach of current observations. One complication is that, scintillation in the interstellar medium and intergalactic medium along the line of sight will introduce additional phase shifts that can be chaotic. Nevertheless, some works suggest that the lensing signal can be reasonably decoupled with scintillation using specific analysis techniques \citep[e.g.,][]{Eichler2017, Katz2020}.

So far, we have only discussed gravitational lensing of a single burst. Obviously, repeating FRB sources can get lensed as well. The actual detectable rates of lensed nonrepeating and repeating FRB sources can be different as their time spans are significantly different and their luminosity functions also appear to differ \citep[e.g.,][]{2020MNRAS.494.2886H}. In general, it should be easier to determine whether the FRB is lensed for repeating FRB sources where all the bursts will be delayed and magnified by the same amount. However, it may not be as straightforward when the separations of the multiple bursts are comparable to the time delays.

\section{Summary}
\label{sec:conclusion}

In this work, we study the strong lensing effects of a point-mass lens model and a point-mass $+$ external shear lens model on single-peak FRBs. Motivated by discoveries of double-peaked FRB sources, we particularly focus on doubly imaged cases. We find that the point-mass lens model can produce two peaks, and the leading peak will be always brighter (more magnified) than the trailing peak. The point-mass $+$ external shear lens model can produce up to four images. Considering only two-image configurations, the point-mass $+$ external shear lens model is capable of producing a brighter trailing peak.

For a point-mass lens model, the lens redshift $z_l$ can be inferred when the opening angle, in addition to the leading-to-trailing flux ratio $R$ and time delay $\Delta t$, of the two lensed peaks can be measured. Even if the two peaks cannot be spatially resolved, a lower limit in $z_{l}$ can still be obtained based on the angular resolution of the observations. In addition, the combination of lens mass $M$ and redshift $z_l$ in the form of $M(1+z_l)$ can be directly computed from the observed $R$ and $\Delta t$ (Equation~\ref{eq:mass-z}). As an example, we consider lensing-induced double peaks with $R=1.95$ and $\Delta t=11$ ms, which are the observed values of one double-peaked FRB source---FRB 130729---and find $M = 828/(1+z_{l})M_{\odot}$.

For a point-mass $+$ external shear lens model, there is not a one-to-one correspondence between $M(1+z_l)$ and $R$ and $\Delta t$ due to the extra freedom from the external shear. Nevertheless, we show that constraints in $M(1+z_l)$ can still be obtained from $R$ and $\Delta t$ for a given external shear strength $\gamma$ (Section~\ref{sec:pointshear}). For the same lensing-induced double peaks with $R=1.95$ and $\Delta t=11$ ms discussed above, the possible ranges of $M(1+z_l)$ for a moderate external shear of $\gamma=0.01$ are $[713 M_\odot, 1089 M_\odot]$ and $[3456 M_\odot, 13953 M_\odot]$. Considering another example with $R=0.91$ and $\Delta t=2.4$ ms, which are the observed values for another double-peaked FRB source---FRB 121002---the range of $M(1+z_l)$ for a moderate external shear of $\gamma=0.01$ is $[505 M_\odot, 3044M_\odot]$. Moreover, we find that tighter constraints in $M(1+z_l)$ can be obtained when the observed $R$ is large.

To date, only a few double-peaked FRB sources have been discovered, none of which can be firmly associated with strong lensing. Nevertheless, given the high event rate and growing efforts in FRB searches \citep[e.g.,][]{Zhang18, CHIME2019,2019ApJ...885L..24C, Cho2020, Zhang20, Zhu2020}, FRB sources strongly lensed by point-mass lenses are expected to be discovered in the near future. Due to the unpredictability and short-lived nature of FRBs, it may be impossible to perform a full lens modeling with the discovery data. The method presented in this work provides an alternative, straightforward way of constraining the lens mass from easily obtained observables of $R$ and $\Delta t$.

\section{Acknowledgements}
The authors thank the anonymous referee for helpful comments and thank Xuefeng Wu, Jun Zhang, and Zuhui Fan for helpful discussions and suggestions. This work is supported by the NSFC (No. 11673065, U1931210, 11273061). We acknowledge the cosmology simulation database (CSD) in the National Basic Science Data Center (NBSDC) and its funds the Task of the CAS (No. 2020000088). Yiping Shu acknowledges support from the Max Planck Society and the Alexander von Humboldt Foundation in the framework of the Max Planck--Humboldt Research Award endowed by the Federal Ministry of Education and Research.

\end{document}